\documentclass[format=acmsmall]{acmart}\usepackage[]{graphicx}\usepackage[]{color}
\renewcommand\footnotetextcopyrightpermission[1]{}

\AtBeginDocument{%
  \providecommand\BibTeX{{%
    \normalfont B\kern-0.5em{\scshape i\kern-0.25em b}\kern-0.8em\TeX}}}

\setcopyright{none}
\acmDOI{00000/xxxx.xxxx}
\startPage{1}

\acmJournal{CSUR}
\acmVolume{99}
\acmNumber{0}
\acmArticle{0}
\acmMonth{0}

\usepackage{amsthm}
\usepackage{booktabs} 
\usepackage{array}
\usepackage[utf8]{inputenc} 
\usepackage{multicol} %
\usepackage{multirow}
\usepackage{amsmath,wasysym,pifont,algorithm,algpseudocode,xspace}
\usepackage{xfrac}
\usepackage{mathtools}
\usepackage{caption}
\usepackage{subcaption}
\usepackage{paralist}
\usepackage{pifont}

\usepackage{graphicx}
\usepackage{pdflscape} 
\usepackage{cryptocode}

\usepackage[strict]{changepage} 

\DeclareCaptionFont{blue}{\color{LightSteelBlue3}}

\usepackage{adjustbox}
\usepackage{aeguill}

\theoremstyle{plain}

\theoremstyle{definition}

\newcommand{\CASCAde}{European Research Council (ERC) Starting Grant CASCAde (GA n\textsuperscript{o} 716980)}

\begin{document}
\let\\\relax 

\title{A Survey on Hardware Approaches for Remote Attestation in Network Infrastructures}
\author{Ioannis Sfyrakis}
\email{ioannis.sfyrakis@newcastle.ac.uk}
\orcid{0000-0003-2932-9064}
\author{Thomas Gro{\ss}}
\email{thomas.gross@newcastle.ac.uk}
\affiliation{%
  \institution{School of Computing, Newcastle University}
  \streetaddress{1 Science Square, Science Central}
  \city{Newcastle upon Tyne}
  \postcode{NE4 5TG}
}

\renewcommand{\shortauthors}{Sfyrakis and Gro{\ss}}

\begin{abstract}
Remote attestation schemes have been utilized for assuring the integrity of a network node to a remote verifier.
In recent years, a number of remote attestation schemes have been proposed for various contexts such as cloud computing, Internet of Things (IoTs) and critical network infrastructures. These attestation schemes provide a different perspective in terms of security objectives, scalability and efficiency. In this report, we focus on remote attestation schemes that use a hardware device and cryptographic primitives to assist with the attestation of nodes in a network infrastructure. We also point towards the open research challenges that await the research community and propose possible avenues of addressing these challenges.
\end{abstract}

\fancyfoot{}
\maketitle
\thispagestyle{empty}

\section{Introduction}
\label{sec:intro}
Network Infrastructures (NIs) exist in companies, industry and critical infrastructure sectors. In the context of companies, the network infrastructure is considered as the backbone of the operation and offering services to customers. For instance, in cloud computing used by many of today's companies, cloud network infrastructures provide the base service model to realize cloud services. Another area where we can find network infrastructures is in any of the 16 critical infrastructures sectors~\cite{cisa:2020} such as energy and financial services. Both application areas of network infrastructures depend on the integrity of their nodes to provide a trustworthy service to clients.

Compromising the integrity of nodes in a network infrastructure such as virtual network devices in Infrastructure-as-a-Service (IaaS) clouds~\cite{bays:2015} or even sensor nodes in an energy smart grid~\cite{enisa:smart_grid:2012, sakhini:iot_smartgrid:2019} can have tremendous effect in the trustworthiness of these systems. Attacks performed in these network infrastructures~\cite{chasaki:attacks_networkinfra:2011} can have a devastating effect, which can affect industry and society~\cite{cisa:alert:2018}.

Current network infrastructures in cloud computing and critical infrastructure systems have a large number of nodes which makes attestation not feasible when attesting only one node at a time. Remote attestation schemes traditionally focus on attesting the integrity of a single node and providing evidence for its current state to a remote verifier. There are proposals that attempt to tackle the issue of attesting more than one node especially in the realm of wireless sensor networks~\cite{steiner:2016}.

The security objectives and assumptions in these remote attestation schemes are different where a number of attestation schemes provide protection from software-based attacks and/or physical attacks. Other attestation schemes protect from Denial of Service (DoS) attacks~\cite{rasmussen2016DAC}.

\textbf{Our contribution:}
In this paper, we offer an overview on the current state of the art regarding hardware-based architectures for attestation and their related attestation schemes in network infrastructures that use cryptography. We focus on attestation schemes that rely on hardware to prove that the software of a network node was not tampered with. In addition, we review attestation schemes that depend on cryptographic primitives and protocols to prove that the network node is in a legitimate state. Architectures and attestation schemes that do not meet these criteria are not included in this paper.  

\subsection{Outline}
\label{subsec:outline}
In section~\ref{sec:background} we discuss background information regarding network infrastructures and remote attestation. Section~\ref{sec:arch-ra} provides an overview of the main hardware based architectures that can realize remote attestation schemes. Section~\ref{sec:hw-ra} continues with a discussion on the hardware-based attestation schemes. Section~\ref{sec:hy-ra} focuses on the remote attestation schemes that are categorized as hybrid attestation schemes. Section~\ref{sec:open-res} discusses the open research problems and possible ways of resolving them. 

%

\begin{table}[]
\centering
\caption{Notation Summary}
\label{tab:notation-summary}
\begin{tabular}{@{}ll@{}}
\toprule
Term & Description \\ \midrule
 $V$    & Verifier \\
 $P$    & Prover   \\
 $n_V$  & Challenge nonce from $V$ \\ 
 $\sigma$ & Signature of attestation \\
 $s$ & Internal state \\
 $k$ & Security parameter \\
 $\mathcal{S}$ & Sign operation \\
 $\mathcal{H}$ & Hash operation \\
 \bottomrule
\end{tabular}%
\end{table}

\section{Background}
\label{sec:background}

\subsection{Network Infrastructures}
\begin{figure}[t]
\begin{center}
\includegraphics[scale=0.4]{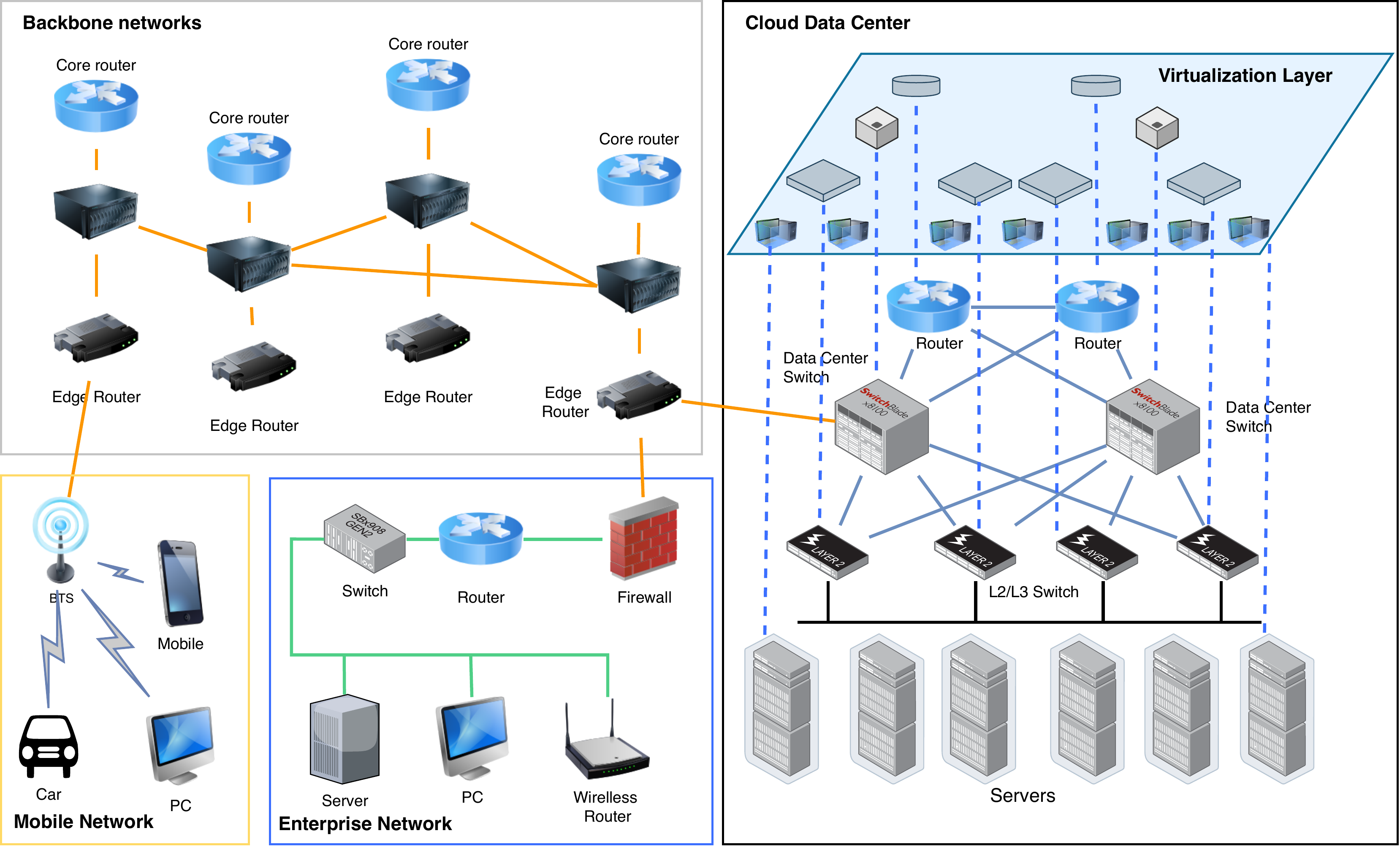}
\caption{Network infrastructures deployed in various contexts such as cloud computing, enterprise networks, mobile networks and backbone networks}\label{fig:net-infra}
\end{center}
\end{figure}

Network Infrastructures (NI) provide the telecommunication backbone for a variety of contexts such as cloud computing, enterprise networks and critical infrastructures. NIs consist of a connected group of computer systems linked with a number of network hardware components. As seen in Figure~\ref{fig:net-infra} the main hardware components are routers, wireless access points, switches and firewalls. Each hardware component employs a software layer that is managed by network administrators. The software layer can be as simple as firmware used in routers or switches to full blown operating systems to manage cloud infrastructure networking.  

NIs are categorized as either open or closed. Open NIs means that the infrastructure employs an open architecture such as the Internet. In a closed NI the network is confined to a certain boundaries. For instance, an enterprise intranet is considered a closed NI. Another categorization of NIs is the method the nodes in the network use to communicate with their peers. The nodes in the network can either communicate via wired network connection or a wireless one or even a combination of the two methods.    

In recent years there has been many research works regarding NIs to create higher-level abstractions. One method is to decouple the network functions from network devices using virtualization, which is called network function virtualization (NFV)~\cite{mijumbi:nfv_survey:2015}. For instance a network function such as a firewall can be deployed as virtual network function in software which enhances its programmability. Another method is by using Software Defined Networking (SDN)~\cite{kreutz:sdn_survey:2014}. SDN separates the network control from the actual network devices. This means that the network control becomes programmable via open interfaces such as OpenFlow~\cite{lara:openflow_survey:2014}. In this setting the network devices become forwarding devices and a centralized controller implements the control logic. 

As we have seen so far there is a variety of technologies that are created on top of NIs. Even though the network devices are used in a different way depending on which method is selected, we still require that the devices are protected from malicious users. This can be achieved by using remote attestation schemes for NIs to ensure that nodes are authenticated and have not been compromised by a malicious user.  

\subsection{Remote Attestation}
\label{sec:remote}

Remote attestation is a mechanism that enables a remote verifier to ascertain the integrity of a host that it is in a known state. This mechanism follows a challenge-response protocol and the main entities in this protocol are the \textit{Prover}($P$) and the \textit{Verifier}($V$). The goal of an honest, non-compromised prover is to construct a response such that it can convince a \textit{Verifier} that the prover is in an acceptable state for the time the attestation was requested.

There are three main types of remote attestation according to the taxonomy by Steiner and Lupu~\cite{steiner:2016}: \begin{enumerate}
	\item \textit{software-based attestation} where the attestation mechanism relies on strict time measurements to convince a verifier that the prover is in a legitimate state. 
	\item \textit{hardware-based attestation} where a tamper-resistant hardware component is used to execute the attestation routines in a secure environment. 
	\item \textit{hybrid attestation} where the attestation method uses hardware/software co-design to offer attestation services but does not require complex hardware components. In hybrid attestation the main requirements are minimal hardware components such as Read Only Memory (ROM) and a Memory Protection Unit (MPU) that a device with limited capabilities can include in its design.
\end{enumerate} 

\begin{figure}[t]
\begin{center}
\includegraphics[scale=0.7]{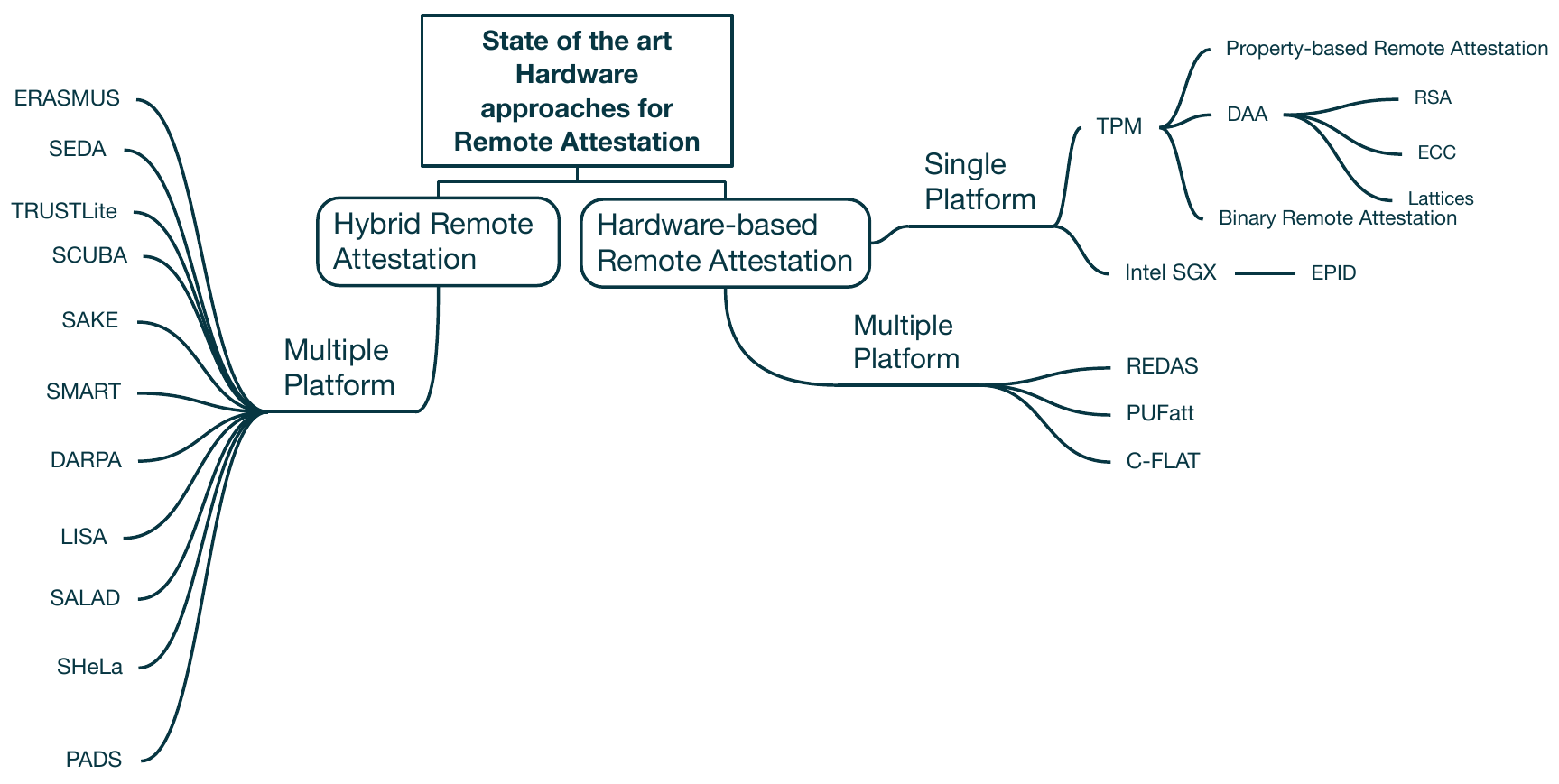}
\caption{Taxonomy of state-of-the-art hardware approaches for remote attestation}
\label{fig:tax-hw}
\end{center}
\end{figure}

In this study, we aim to review the state-of-the-art remote attestation approaches that are either based on hardware components or they offer a software/hardware co-design architecture that realizes the attestation scheme. In addition, we focus on remote attestation schemes that utilize cryptographic primitives as part of their attestation protocol. We further categorize the remote attestation schemes in terms of the number of attesting nodes in a network infrastructure. The first category attests one node each time while the second category the attestation happens in groups of nodes. Strong security is of paramount importance in today's environment where attackers are able to launch attacks to network infrastructures. Therefore, cryptography can be used in a way that benefits the attestation of nodes in a network infrastructure.

Figure~\ref{fig:tax-hw} depicts the taxonomy of the hardware approaches for remote attestation in network infrastructures. We structure this report around the taxonomy and discuss each category for the remote attestation schemes. In addition we compare the research works in each category in terms of how the attestation works in network infrastructures and the number of attested nodes the method supports. The cryptographic primitives used in each attestation research works are compared and discuss the reasoning for adding them to each attestation scheme.

We acknowledge previous works in remote attestation that provide a generic protocol view for attestation schemes ~\cite{francillon2014minimalist,coker:principles_ra:2011,datta:logic_trustedcomputing:2009, armknecht:software_attestation:2013, li:framework_softwareattestation:2014} and we adopt that the attestation involves three main stages: (1) \texttt{Challenge}, (2) \texttt{Attest} and (3) \texttt{Verify}. 

\begin{figure}[t]
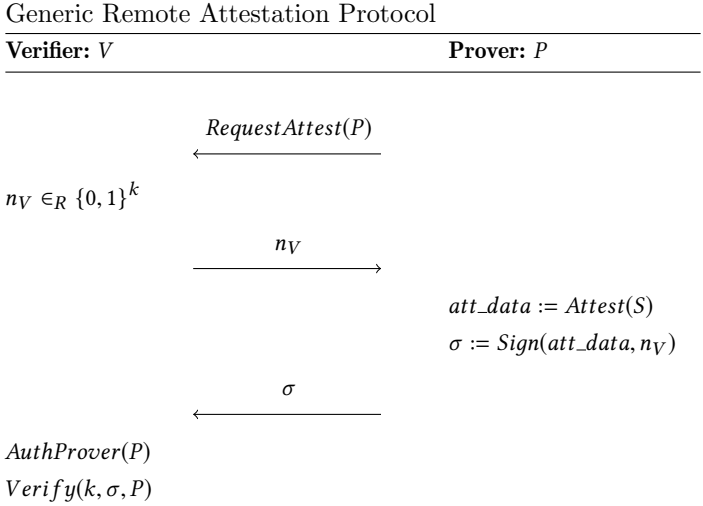

\begin{center}
\procedure[colsep=1em]{Generic Remote Attestation Protocol} {
	\textbf{Verifier: $V$} \< \< \textbf{Prover: $P$} \\[0.1\baselineskip][\hline]	\\
	\< \sendmessageleft*[2.5cm]{RequestAttest(P)} \< \\
	\text{$n_V \in_R \{0,1\}^k$}  \< \< \\
	\< \sendmessageright*[2.5cm]{n_V} \< \\
	\< \<  \text{$att\_data:= Attest(S)$} \\
	\< \<  \text{$\sigma:=Sign(att\_data,n_V)$} \\
	\< \sendmessageleft*[2.5cm]{\sigma} \< \\
	\text{$AuthProver(P)$} \< \< \\
	\text{$Verify(k,\sigma, P)$} \< \< \\
		}
	\caption{Generic Remote Attestation Protocol. In this protocol we assume that the verifier and prover have generated key pairs according to parameter $k$.}
\label{tab:genattprot}
\end{center}	
\end{figure}

Figure~\ref{tab:genattprot} illustrates a concise overview of a generic remote attestation protocol that uses cryptography. The attestation protocol is similar to the ones in~\cite{francillon2014minimalist,coker:principles_ra:2011, datta:logic_trustedcomputing:2009}. In this generic protocol view of remote attestation we assume that both the \textit{Verifier} and the \textit{Prover} have generated key pairs according to a parameter $k$. The \textit{Prover} initiates the protocol by sending a message to the \textit{Verifier} requesting for an attestation to be performed in her platform. Then the \textit{Verifier} responds by first generating a nonce that is used a challenge nonce and sending the nonce back to the \textit{Prover}. The \textit{Prover} upon receiving the challenge nonce fro the \textit{Verifier} then she performs the attestation process \texttt{Attest} and generating the appropriate attestation data as evidence for the \textit{Verifier} to prove. For instance, in the TPM one might use as attestation evidence the current contents of the Platform Configuration Registers (PCRs) which demonstrate the current state of the platform. After the attestation data has been collected then the \textit{Prover} signs the attestation data and the challenge nonce using a signature scheme. The signature $\sigma$ is then communicated to the \textit{Verifier}. In this instance, the \textit{Verifier} is starting the authentication process \texttt{AuthProver(P)} to authenticate that the \textit{Verifier} is talking to the right \textit{Prover}. The next step in the remote attestation protocol is to verify the signature received from the \textit{Prover} using the \texttt{Verify(k,$\sigma$,$P$)} process. The outcome of the verification process is either true or false.    

\section{Hardware-based Architectures for Remote Attestation}
\label{sec:arch-ra}

In this section, we review the main architectures based on hardware device or component for realizing remote attestation schemes. Table~\ref{tab:hardware-attestation-summary} provides an overview of all the architectures that are used for realizing remote attestation schemes. 

\subsection{TPM}
One of the most prevalent examples of a discrete tamper-resistant hardware component for remote attestation is the Trusted Platform Module (TPM). This device is added in motherboards as a way to provide trusted computing facilities to the platform. Currently there are two versions of the TPM. The first version of the TPM v1.2 ~\cite{tpm1.2} is based on RSA cryptography and the second version TPM v2.0 is based on elliptic curve cryptography~\cite{brickell2004direct,tcg:tpm2}.

The TPM works as coprocessor and is able to store keys and realize remote attestation. The Platform Configuration Registers (PCRs) in the TPM hold hashes of code or data that are used in remote attestation. The published specification for TPM v1.2 outlines the functionality and cryptographic primitives that are required for the hardware component. Generating random data in a TPM is supported using a Randomness Number Generator (RNG). The main cryptographic primitive that is supported is RSA key generation and signature schemes. For calculating the hashes the TPM uses the SHA-1 hash algorithm, which is now considered vulnerable to attacks. The first key enclosed in the TPM is the Endorsement Key (EK) and is generated during the manufacturing process. This key is used in almost all key related operations in the TPM. Other keys that can be generated are the Attestation Identity Keys (AIK) used in digital signature operations and the storage keys used in encryption and decryption of data.

The first approach for supporting remote attestation in TPM v1.2 used a Trusted Third Party (TTP) called Privacy-CA~\cite{chen:priv_ca:2010} that raised privacy concerns. Another remote attestation scheme was added that provides privacy-preserving remote attestation based on a group signature scheme~\cite{brickell2004direct} based on RSA called Direct Anonymous Attestation (DAA).  

The TPM specification was updated to version 2.0~\cite{tcg:tpm2:architecture} provides a number of new features for the trusted hardware component. First, the TPM v2.0 supports a larger variety of cryptographic algorithms, and especially for privacy-preserving attestation the implementation is less rigid giving room for realizing different flavors of DAA-like schemes. Second, this version of the TPM supports multiple banks of PCRs and three key hierarchies. 

One of the main disadvantages of using TPMs in network infrastructures is that they support high-power hardware. They are used in hardware platforms that have enough processing power to assist with the computation of attestation schemes. Another disadvantage that exists in TPMs is that they privacy-preserving attestation schemes are considered complex and they are difficult to deploy in large network infrastructures.

\begin{table}[t]
\centering
\caption{Overview of main hardware-based architectures that support attestation schemes}
\label{tab:hardware-attestation-summary}
\resizebox{\textwidth}{!}{%
\begin{tabular}{@{}llllllll@{}}
\toprule
\textbf{Architecture} &
\textbf{Approach} &
\textbf{Features} &
\textbf{Advantages} &
\textbf{Disadvantages} &
  \begin{tabular}[c]{@{}l@{}}\textbf{Cryptography}\\ \textbf{Support} \end{tabular} &
  \begin{tabular}[c]{@{}l@{}} \textbf{Theoretical} \\\textbf{Results} \end{tabular} &
\textbf{Examples} \\ \midrule
\textbf{TPM}~\cite{tpm1.2, tcg:tpm2} &
  co-processor &
  \begin{tabular}[c]{@{}l@{}}Secure key storage, \\ Secure memory, \\
  Data sealing, \\ Attest software configurations, \\ Bind keys to TPM device  \end{tabular} &
  \begin{tabular}[c]{@{}l@{}}Attestation algorithms in HW, \\ Algorithmic agility (v2.0), \\ Support privacy-preserving \\ attestation \end{tabular} &
  \begin{tabular}[c]{@{}l@{}} Req. high-power HW, \\ Low scalability, \\ Complex privacy-preserving \\ protocols \end{tabular}   &
  \begin{tabular}[c]{@{}l@{}}RSA, ECC, \\ HMAC\end{tabular} &
  \begin{tabular}[c]{@{}l@{}}DAA,\\  Privacy-CA\end{tabular} &
   ~\cite{brickell2004direct,chen:priv_ca:2010,tan:tpm_wsn:2015} \\ \\[2pt]
\textbf{SGX}~\cite{mckeen:intel_sgxra:2013, anati:intel_sgxra:2013, johnson:sgx_epid_prov_attest:2016} &
 \begin{tabular}[c]{@{}l@{}} Special CPU \\ instructions\end{tabular} &
  \begin{tabular}[c]{@{}l@{}}Protected software in enclave, \\ Sealing \\ Local/Remote attestation, \\ \end{tabular} &
  \begin{tabular}[c]{@{}l@{}}Support privacy-preserving \\ attestation\end{tabular} &
  \begin{tabular}[c]{@{}l@{}}Req. specific CPUs, \\ Closed ecosystem for \\ attestation services\end{tabular} &
  ECC, CMAC &
  EPID &
 ~\cite{wang:intel_sgx_ra:2017, chen:sgx_opera_ra:2019}  \\ \\[2pt]
\textbf{SMART}~\cite{defrawy:2012,francillon2014minimalist} &
  \begin{tabular}[c]{@{}l@{}} HW/SW \\ co-design \end{tabular} &
 \begin{tabular}[c]{@{}l@{}}Read-only verification code,\\ Secure key storage,\\ Attestation ROM \\ Atomic execution\end{tabular} & 
  \begin{tabular}[c]{@{}l@{}}Minimal HW changes, \\ Supports low-power \\  devices\end{tabular} &
  \begin{tabular}[c]{@{}l@{}} Static attestation \\ key/code, \\ One attestation instance, \\ Limited crypto 
  support\end{tabular} &
   HMAC &
   None &
 \begin{tabular}[c]{@{}l@{}}~\cite{ibrahim:seed:2017,carpent:lisa_ra:2017} \\ ~\cite{ibrahim:us-aid:2018,carpent:self_measurement_ra:2019}\end{tabular}  \\ \\[2pt]
\textbf{TrustLite}~\cite{koeberl:trustlite:2014} &
  \begin{tabular}[c]{@{}l@{}}HW/SW \\ co-design\end{tabular} &
  \begin{tabular}[c]{@{}l@{}}Execution-Aware Memory \\ Protection Unit, \\ Local attestation \\ between trustlets, \\ Secure Loader, \\ Trusted communication \\ between trustlets\end{tabular} &
  \begin{tabular}[c]{@{}l@{}}Minimal HW changes, \\ Updateable attestation code \\ and security policies, \\ Support low-power devices\end{tabular}  &
  \begin{tabular}[c]{@{}l@{}}Support only \\ local attestation, \\ Bespoke OS, \\ Limited crypto support,  \\ Static software configuration \end{tabular} &
\begin{tabular}[c]{@{}l@{}}Cryptographic \\ Hash\end{tabular}   &
  None &
 ~\cite{ibrahim:seed:2017,ibrahim:us-aid:2018,ambrosin:pads:2018}  \\ \\[2pt]
\textbf{Tytan}~\cite{braser:tytan:2015} &
  \begin{tabular}[c]{@{}l@{}}HW/SW \\ co-design\end{tabular} &
  \begin{tabular}[c]{@{}l@{}}Hardware-assisted dynamic \\ root of trust, \\ Secure IPC, \\ Local/remote attestation, \\ Secure boot, Platform key\end{tabular} &
\begin{tabular}[c]{@{}l@{}}Support low-power devices, \\ Dynamic configuration \\ of access control rules, \\ Local attestation of \\ multiple trustlets\end{tabular}   &
 \begin{tabular}[c]{@{}l@{}}Update tasks at runtime, \\ Limited crypto support \\\end{tabular}  &
  \begin{tabular}[c]{@{}l@{}}Cryptographic \\ Hash\end{tabular} &
  None &
~\cite{ambrosin:sana:2016}   \\ \\[2pt]
\textbf{Sanctum}~\cite{costan:sanctum:2016} &
  \begin{tabular}[c]{@{}l@{}}HW/SW \\ co-design\end{tabular} &
  \begin{tabular}[c]{@{}l@{}}Security monitor for enclaves,  \\ Local attestation \end{tabular}  &
\begin{tabular}[c]{@{}l@{}}Minimal/minimal invasive \\ HW changes, \\ Support low-power devices, \\ Attestation chain of trust \end{tabular}   &
\begin{tabular}[c]{@{}l@{}}Req. bespoke SW, \\ Synchronous attestation \\ Relies on signing \\ enclave for attestation\end{tabular}   &
 \begin{tabular}[c]{@{}l@{}}Cryptographic \\ Hash, \\ Symmetric `encryption, \\ RSA\end{tabular}   &
 None  &
 ~\cite{lebedev:sanctum_ra:2018}  \\ \bottomrule
\end{tabular}%
}
\end{table}

\subsection{Intel SGX}
Intel Software Guard Extensions (SGX)~\cite{mckeen:intel_sgxra:2013} is a trusted execution environment that was first announced in 2013 and supports both local and remote attestation. Applications can support Intel SGX by partitioning the application in two parts the untrusted and trusted part. We create an enclave with the trusted part of the application. Intel SGX assures the confidentiality and integrity of code and data that reside in an enclave. 
In addition, there is no other software such as OS or a hypervisor that can access the memory region that the enclave resides. The enclave pages reside in a memory region called Processor Reserved Memory (PRM). This region includes two structures for the enclaves, the Enclave Page Cache (EPC) and Enclave Page Cache Map (EPCM). The EPC store the code and data of the enclave while the EPCM holds state information for the enclave. 

Untrusted software create and initialize the enclave and the hardware assures that the enclave can only be modified before it is initialized. During the initialization phase all the contents of the enclave such as its code and data are measured and this measurement can become the basis for future local and remote attestations. 

One feature of Intel SGX is that an enclave can attest to another that it has been loaded correctly by producing a report which includes information about the author of the enclave or the enclave measurement. This process consists the method for local attestation. Intel SGX uses a specific enclave called quoting enclave that wraps the report into a quote. More precisely the quote is signed using an asymmetric attestation key. This process is part of the Enhanced Privacy Identifier (EPID)~\cite{brickell2007enhanced} signature scheme provided by Intel, which is an extension of the DAA scheme used in TPMs. A remote verifier can then receive the quote and start the verification process with a corresponding verification key that is provided by Intel.

The attestation in Intel SGX requires specific CPUs and provides a closed ecosystem for realizing attestation schemes that are totally controlled by Intel. There are efforts that attempt to create third party attestation services~\cite{scarlata:sgx_attest_prim:2018} using Intel SGX but it is still required to use an Intel provided enclave named Provisioning Certification Enclave (PCE) that acts as a local certificate authority.  

\subsection{SMART}
SMART~\cite{defrawy:2012} is a reference hardware architecture that supports attestation in low-powered devices. The main features of this architecture is that it includes immutable attestation code stored in a Read-only Memory (ROM) that guarantees that the code is not changed after it is placed there. In addition, SMART supports secure key storage and atomic execution of operations. 

The design principles behind this architecture propose that changes to hardware components and interfaces should be kept to a minimum. As the attestation code is immutable SMART architecture can support only one attestation scheme instance. 

\subsection{TrustLite}
The TrustLite~\cite{koeberl:trustlite:2014} hardware architecture is a flexible and efficient software isolation platform for low-powered devices. This architecture introduces trustlets, which are trusted tasks developed to realize an application. The goal of trustlets is to isolate the enclosed software components and protect the confidentiality and integrity of their code and data. The memory protection scheme in this architecture involves the Execution-Aware Memory Protection Unit (EA-MPU). The EA-MPU can be programmed in software and offers a flexible allocation of memory across memory and I/O components. Another feature of the architecture is the Secure Loader, which loads the selected trustlets alongside their data into memory. Another goal of the Secure Loader is to program the Memory Protection Unit (MPU) to protect its own code and the memory regions of all trustlets. The Secure Loader also start the chain of trust process for remote attestation.

One of the advantages of this architecture is that it enables the local attestation between multiple trustlets. Trusted communication between trustlets happens using a simple handshake protocol between a Trustlet A and Trustlet B. Trustlet A is the initiator that verifies the platform configuration and the security policy of the MPU for Trustlet B who is the responder. The initiator can also verify the cryptographic hash of the responder's internal state so that it can be assured that is not modified. Then the responder can also perform a local attestation on the initiator using the same attestation process as the initiator. A cryptographic hash over the identifiers of the trustlets and their corresponding nonces comprise the cryptographic session token. The token can be used to authenticate bidirectional messages between the two trustlets. 

Another advantage of TrustLite is that its security extensions do not depend on the CPU instruction set and can be adjusted using software. Thus all the related software such as the Secure Loader and security policies can be updated. This is in contrast with the SMART architecture where the attestation code and key are not updateable.  

One of the main disadvantages of this security architecture is that it only support the local attestation of trustlets that remain on the same system. Another disadvantage is that TrustLite requires a custom built software in order to support the memory protection scheme and the user tasks. In terms of cryptography support it only uses a cryptographic hash for the attestation process. Of course the architecture could be extended with additional cryptographic components such as cryptographic accelerators. The default configuration is to just enable a cryptographic hash for measuring the state of a trustlet.  In addition, the static software configuration does not enable to change the software configuration dynamically during runtime. 

\subsection{Tytan}
Tytan ~\cite{braser:tytan:2015} is a hardware security architecture for low-powered devices. The system architecture of Tytan consists of ten different components. Tasks represent the applications in embedded systems and there are two types of tasks. The first type is the normal tasks that are isolated from other tasks but not from the OS. The second type is the secure tasks that are isolated from all the other software. In addition, each task has a unique identifier. 

The Execution-aware Memory Protection Unit (EA-MPU) hardware component enforces memory access control and makes sure that each task can only access the memory it has been allocated. The hardware component is also coupled with a driver that enables the dynamic handling of tasks and the loading or unloading of a secure task. Tytan is assigned a platform key that the EA-MPU enforces the access to the key. 

Tytan architecture includes a real time OS to provide a real-time scheduling.  
The main features of this architecture include the strong isolation of dynamically configured software components using dynamic root of trust and real time guarantees. Another feature of Tytan is the local and remote attestation. Tytan also supports secure inter-process communication (IPC) between two parties using authentication. 

Attestation in Tytan is realized by executing the Root of Trust of Measurement  (RTM) task. This task is executed to prove the integrity of a either a local or a remote by computing a cryptographic hash over the binary code of each attested task. In order to prove authenticity to a local verifier the task identifier is used. For remote attestation the a Message Authentication Code (MAC) is computed alongside an attestation key.   

In terms of disadvantages this architecture does not support the updating of tasks during runtime and only supports a cryptographic hash method for the attestation process.  

\subsection{Sanctum}
Sanctum~\cite{costan:sanctum:2016,lebedev:sanctum_ra:2018} is an architecture that depends on minimal and minimal invasive changes to the hardware. It uses a trusted software component called security monitor to yield security properties similar to Intel SGX. The architecture relies on a hardware/software co-design to isolate and protect the integrity of applications. The authors make slight adjustments to well understood building blocks and mechanisms. The CPU building blocks do not require any adjustments as Sanctum only makes changes to the interfaces of these building blocks. Thus, the integration of Sanctum to other hardware platforms is easier to achieve in comparison to Intel SGX which requires specific CPUs in order to function. 

Sanctum relies on a select set of trusted software components to drive similar security mechanisms with those in Intel SGX. For instance both systems support a method to locally attest the integrity of software. Sanctum provides an open-source implementation of the security monitor which is portable to different CPU architectures and amenable to rigorous security analysis. In contrast, the Intel SGX relies on microcode which its source code is not openly available and is specific to the CPU models.  

The trusted software components replace the microcode that Intel SGX uses. There are three software components that are part of the Sanctum architecture. The first component is the measurement root that is burned in an on-chip ROM. The goal of the measurement root is to execute successfully three tasks: compute a cryptographic hash from the security monitor which is added as input to both generate a monitor attestation key pair and attestation certificate.

The second component is the security monitor which takes over when the measurement root has managed to construct the attestation measurement chain. The security monitor offers an API that focuses on managing enclaves and allocating memory regions. The API can be invoked both from the OS and the enclaves. Timing attacks can deployed in such a setting for the API. For this reason the security monitor is prohibited from accessing any enclave data and from accessing any memory regions related to the attestation key. In addition, the security monitor is not allowed to perform any cryptographic operations that use attestation keys. Thus, computing attestation signatures is deferred to the signing enclave.

The third component is the signing enclave which executes the signing algorithm. The reasoning behind this design decision is to avoid any timing attacks and remove the dependency on the security monitor to compute the signatures for attestation. The signing enclave is executed inside the security monitor as an enclave isolated from other software. The security monitor evaluates any calls to the signing enclave by checking the known measurement of the signing enclave. If the call is successfully evaluated, then it proceeds to the singing enclave. For instance, the first call that the signing enclave receives is the monitor's private attestation key which is then copied to the enclave's memory. The signing enclave can authenticate another enclave and uses a mechanism similar to the Intel SGX's reporting mechanism called mailbox. 

Mailboxes are part of the enclave's metadata and are specified during construction. This is a synchronous messaging mechanism that requires both sending and receiving enclave to run simultaneously and exist in memory so that they can communicate. The receiving enclave uses an API call that includes the mailbox that will receive the message and the sending enclave's identity. The sending enclave which want to be authenticated sends an API call to the receiving enclave with the identity of the receiving enclave and the mailbox inside the receiving enclave. Then the receiving enclave issues a read API call to request the contents of the message to be moved from the mailbox to the enclave's memory. The receiving enclave is assured of the sending enclave's identity when the API call to read is successful.

\section{Hardware-based attestation} 
\label{sec:hw-ra}

In this section we focus on the attestation methods that use a tamper-resistant hardware component to drive the attestation process. We separate the methods used in two categories depending on if the methods support attesting only one node each time or a group of nodes.  We also discuss two variations of the hardware-based approach. First, a discrete tamper-resistant hardware component embedded in a motherboard provides the required attestation services such as generating and storing cryptographic keys, signing attested data. Second, there are co-processor hardware components that assist the attestation process and collaborate with other hardware components.

\subsection{Single Platform Attestation}
In this subsection, we focus on approaches that attest one node at a time in a network infrastructure. 

\subsubsection{Trusted Platform Module (TPM)}


Attesting the integrity of a single platform is the focus of most attestation schemes developed during the last 20 years of research. It is a popular problem domain, which has been thoroughly researched and especially for trusted computing. In the following paragraphs we discuss the main schemes that have been developed. 

The protocol in the Trusted Computing Group (TCG) specification for the TPM v1.2 is based on the third party \textit{Privacy Certificate Authority (CA)}~\cite{tcg:2005} to complete the attestation of the platform identify.
When a user purchases a trusted computing platform it does not have an identity. In order for the platform to gain an identity the TPM owner requests a certificate from a trusted third-party Privacy CA. The output of this process is the Attestation Identity Key (AIK) certificate that the platform can use to attest the integrity of the platform. The main disadvantage of this process is that the Privacy CA knows the identity of the platform. If the verifier colludes with the Privacy CA then the integrity of the platform cannot be assured. Another disadvantage of this approach is that the Privacy-CA is a single point of failure and it can become a performance bottleneck if the verifier queries the Privacy-CA for each verification request. Paired dictate this is very good approach regarding this at the station skin skin scheme. The next idea that is included in this report discusses the main aspects off the security and privacy issues what are eyes doing the execution off the other stations game receipt

%


\subsubsection{Direct Anonymous Attestation}

The \textit{Direct Anonymous Attestation} (DAA)~\cite{brickell2004direct} scheme uses zero-knowledge proofs and CL-signatures to enable the attestation of a platform without compromising the identity of the platform. The DAA scheme includes three entities: platform, issuer and verifier. The role of the issuer is to generate and issue credentials to platforms and consult a revocation list to revoke platforms that are considered compromised. The platform entity in the DAA scheme consists of a host computer and a TPM device. Both components in the platform contribute to create signatures that prove to a verifier that the platform which created the signature was indeed issued a credential by the issuer. As a result the verifier checks the proofs and is not able to identify the platform that created the signature.


The initial version proposed is based on RSA and is considered inefficient and forms part of the TCG specification for TPM v1.2. Subsequent works extend the DAA scheme and use Elliptic Curve Cryptography (ECC) and bilinear maps~\cite{brickell:2008} to further improve efficiency of the scheme. This scheme includes three main entities: issuer, prover and verifier. DAA uses group signatures and zero-knowledge proofs to attest that the remote attestation originates from an authentic TPM without disclosing to the verifier the exact identity of the TPM. The issuer in DAA is provided with an anonymous credential instead of an identity for the TPM. Then the TPM proves its identity to the verifier using zero-knowledge proofs and its pseudonym.

There has been a large body of research over the DAA scheme with improvements suggested and new schemes developed that aim to improve the efficiency of the scheme or integrate DAA in other platforms such as ARM TrustZone. In addition, there is research that manage to extend DAA scheme to encode attributes which can bind multiple TPMs to an anonymous credential.
A recent research thread for direct anonymous attestation is to leverage lattices for creating DAA schemes~\cite{kassem2019daalat} that are post-quantum safe.  

\begin{table*}[t]
\caption{Security properties of DAA schemes.}
\label{tab:properties}
\centering
\scalebox{0.7}{
\begin{tabular}{l c c c c c c c c}
\toprule
\multirow{2}{*}{DAA scheme}    
		    &  &   &  &     &  &   & Security \\
                                     & Unforgeability & Revocation   & Anonymity     &  Unlinkability  & Attributes    & Assumptions   & Model                       \\
\midrule
BCC04~\cite{brickell2004direct}    & $\CIRCLE$    & $\CIRCLE$     & $\CIRCLE$     & $\CIRCLE$     & $\Circle$     & SRSA, DDH       & RO        \\
C06~\cite{camenisch2006aac}    & $\CIRCLE$    & $\Circle$     & $\CIRCLE$     & $\CIRCLE$     & $\Circle$     & SRSA, DDH       & RO        \\
HS07~\cite{hes2007daaemb}               & $\CIRCLE$    & $\CIRCLE$     & $\CIRCLE$     & $\CIRCLE$     & $\CIRCLE$     & SRSA      & RO        \\
BL07~\cite{brickell2007enhanced}      & $\CIRCLE$    & $\CIRCLE$     & $\CIRCLE$     & $\CIRCLE$     & $\Circle$     &    SRSA, DDH             & RO        \\
BCL08~\cite{brickell2008daabm}          & $\CIRCLE$    & $\Circle$     & $\CIRCLE$     & $\CIRCLE$     & $\Circle$     & LRSW, DBDH      & RO        \\
XD08~\cite{xiaofeng2008direct}          & $\CIRCLE$    & $\Circle$     & $\CIRCLE$     & $\CIRCLE$     & $\Circle$     & q-SDH, DDH      & RO        \\
BCL09~\cite{brickell2009daanotions} & $\CIRCLE$    & $\Circle$     & $\CIRCLE$     & $\CIRCLE$     & $\Circle$     & LRSW, DBDH      & RO        \\
BL10a~\cite{brickell2010epidbp}          & $\CIRCLE$    & $\CIRCLE$     & $\CIRCLE$     & $\CIRCLE$     & $\Circle$     & SDH, DDH        & RO        \\
BL10b~\cite{brickell2010daapair}     & $\CIRCLE$    & $\Circle$     & $\CIRCLE$     & $\CIRCLE$     & $\Circle$     & SDH, DDH        & RO        \\
C10a~\cite{chen2010daaless}           & $\CIRCLE$    & $\Circle$     & $\CIRCLE$     & $\CIRCLE$     & $\Circle$     &     q-SDH, DDH            & RO        \\
C10b~\cite{chen2010daabatch}        & $\CIRCLE$    & $\Circle$     & $\CIRCLE$     & $\CIRCLE$     & $\Circle$     & SDH, DDH        & RO        \\
CPS10~\cite{chen2010daaeff}            & $\CIRCLE$    & $\Circle$     & $\CIRCLE$     & $\CIRCLE$     & $\Circle$     & SDH, DDH        & RO        \\
BFG13~\cite{bernhard2013aat}            & $\CIRCLE$    & $\CIRCLE$     & $\CIRCLE$     & $\CIRCLE$     & $\Circle$     & B-LRSW, DDH, CDH       & RO        \\
CU15~\cite{chen2015daaa}                & $\CIRCLE$    & $\Circle$     & $\CIRCLE$     & $\CIRCLE$     & $\CIRCLE$     & LRSW, q-SDH     & RO        \\
CDL16a~\cite{camenisch2016ucdaa}         & $\CIRCLE$    & $\Circle$     & $\CIRCLE$     & $\CIRCLE$     & $\Circle$     &  LRSW, DDH               & UC        \\
CDL16b~\cite{camenisch2016aa}          & $\CIRCLE$    & $\CIRCLE$     & $\CIRCLE$     & $\CIRCLE$     & $\CIRCLE$     & q-SDH           & UC        \\
CCD17~\cite{camenisch2017bind}       & $\CIRCLE$    & $\CIRCLE$     & $\CIRCLE$     & $\CIRCLE$     & $\CIRCLE$     & q-SDH, LRSW     & UC        \\
YCZ18~\cite{yang2018direct}       & $\CIRCLE$    & $\CIRCLE$     & $\CIRCLE$     & $\CIRCLE$     & $\CIRCLE$     & DDH, DBDH, q-SDH    & UC        \\
KCB19~\cite{kassem2019daalat}           & $\CIRCLE$    & $\Circle$     & $\CIRCLE$     & $\CIRCLE$     & $\Circle$     & Ring-ISIS, Ring-LWE  & Standard \\
\bottomrule
\multicolumn{9}{l}{\emph{Note:} $\CIRCLE$: security property supported $\LEFTcircle$: security property partially supported  $\Circle$: security property not supported}
\end{tabular}
}
\end{table*}

\subsubsection{Binary remote attestation} is used to attest the platform integrity. In this scheme the TPM signs the PCR values representing the integrity of the platform. The TPM subsequently sends the TPM signature and the measurement log to the remote verifier. Then the remote verifier is able to attest the current state of the platform. Realization of this scheme is the IBM Integrity Measurement Architecture (IMA)~\cite{Sailer:2004wa} and the trusted linux client (TLC) system~\cite{safford2005trusted}. One of the disadvantages of this scheme is that it discloses the software and hardware configuration of the platform to the remote verifier, which makes the platform vulnerable to malicious attacks. In addition, there is a great variety of software and their configurations  and when the system needs to be updated it makes it difficult to evaluate the configuration integrity of the platform.

\subsubsection{Property-based Attestation (PBA)}
Another scheme used for remote attestation that attempts to overcome the shortcomings of the binary remote attestation is the \textit{Property-based Attestation (PBA)}~\cite{Sadeghi:2004:PAC:1065907.1066038,chen:2008} scheme. In this scheme the platform configurations are mapped to properties which are attested in order to avoid the disclosure of the platform configurations. 


\subsubsection{Physical Unclonable Functions (PUFs)} Another type of hardware component to be used in remote attestation schemes is the Physical Unclonable Functions (PUFs)~\cite{sadeghi2011short,kong:pufatt_ra:2014}. The PUF is a hardware-based security primitive first introduced by Gassend et.al.~\cite{gassend:puf:2002} The authors propose that architecting the construction of the hardware component in a secure way it can protect the attestation process and provide a secure basis for the attestation computations.


\section{Hybrid attestation}
\label{sec:hy-ra}

This attestation method includes a hardware component for the attestation process, but also integrates with the software each network node uses. 
A type of hybrid attestation that is related to our research area is \textit{swarm attestation} or \textit{collective attestation}~\cite{carpent:lisa_ra:2017,asokan2015seda,meng2018dynamic,rabbani2019shela}. The main premise of this hybrid attestation method is the ability to attest a large number of nodes across a network with various interaction patterns.
The main focus is to offer attestation services to remote embedded devices that have limited processing power. Even though collective attestation schemes focus on embedded devices and devices with limited processing power they still have a hardware component that they can rely on to fulfill or assist with the attestation process.

\subsection{Multiple Platform Attestation}
\textit{Swarm attestation} or \textit{collective attestation}~\cite{asokan2015seda,ambrosin2016sana,ibrahim:seed:2017,carpent:lisa_ra:2017,ibrahim2016darpa,kohnhauser2018salad,wedaj2019dads} focuses on remote attestation of multiple platforms at a large scale. This type of attestation is considered a hybrid attestation scheme since it utilizes a software/hardware co-design since the computing platforms are embedded devices with limited capabilities and processing power. In this context a TPM component could not be added to the device. If a TPM component was added to a small embedded device, it would increase production costs and the price of the device.

Collective attestation works by distributing the attestation process across all devices in a network. A spanning tree is created during attestation over the network topology. The root of the spanning tree is the verifier who also initiates the attestation. During the process of the attestation each device validates their child devices and propagate in the spanning tree an aggregate attestation report to their parent devices.

\begin{figure}[t]
	\centering
	\includegraphics[width=0.8\linewidth]{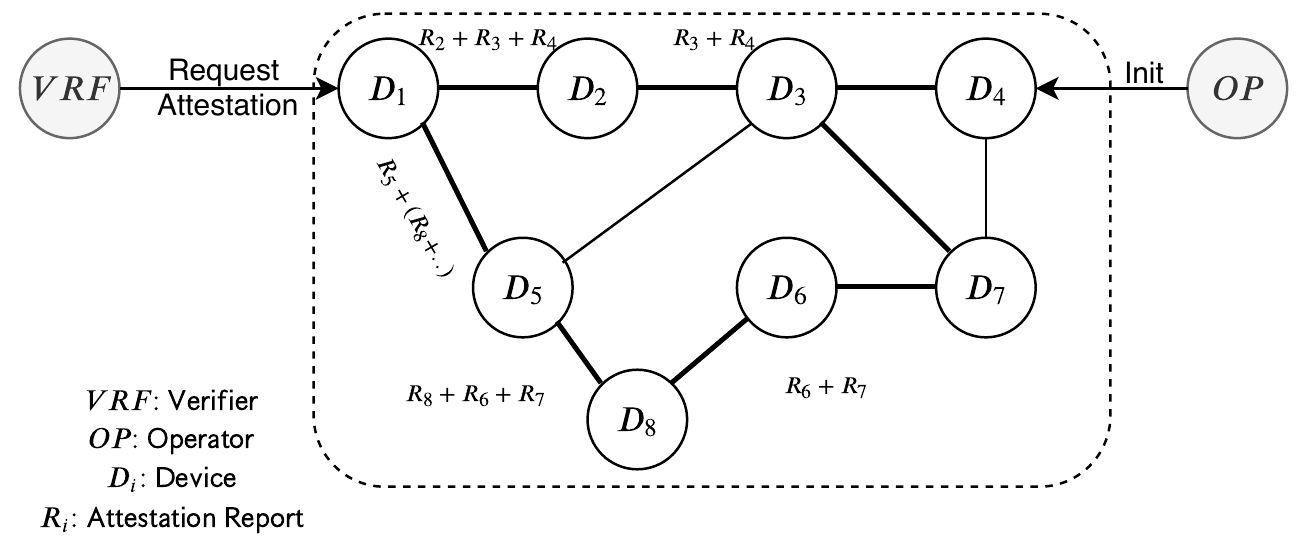}
	\caption{SEDA collective attestation in a swarm of 8 devices~\cite{asokan2015seda}}%
	\label{fig:seda-attestation}
\end{figure}

The first research work to propose a collective attestation scheme was \textit{Scalable Embedded Device Attestation (SEDA)}~\cite{asokan2015seda}. This attestation scheme works by using groups of many provers. SEDA includes an off-line phase where an operator $OP$ initializes the devices in the swarm with a particular software configuration and certificate and an online phase for the attestation process. This attestation scheme requires devices which includes minimal security features and without any complex security hardware such as a TPM. A central verifier $V$ generates attestation requests that are propagated to all nodes of the network topology using a flooding protocol as seen in Figure~\ref{fig:seda-attestation}. Then the verifier accepts the aggregate attestation reports of the devices in the network.

The \textit{Secure Collective Attestation Scheme (SANA)}~\cite{ambrosin2016sana} is a collective attestation scheme for dynamic embedded devices network that employs signature scheme called Optimistic Aggregate Signature scheme (OAS). This scheme integrates aggregate and multi-signatures in which $n$ signers sign messages $m_1, \ldots, m_n$. The majority of the signers sign the default message $M$. Each signature is aggregated into an aggregate signature, which is shorter than signing $n$ different signatures. The main parties of this attestation scheme are: prover ($P$), aggregator ($A$), owner ($O$) and verifier ($V$). A prover $P_i$ constructs the proof of integrity of its software configuration to be sent to the verifier. An aggregator $A_i$ relays the messages between different parties, collects and aggregates the responses from other provers or aggregators. The network owner $O$ is responsible to deploy and maintain the embedded devices that act as provers. A remote verifier $V$ is assured of the software integrity of all provers that are part of the network during the attestation process. SANA also considers the possibility of compromised devices, which do not affect the integrity of the attestation process.

\textit{SeED}~\cite{ibrahim:seed:2017} is another attestation scheme for multiple embedded devices which introduces a non-interactive attestation protocol. The attestation process is executed at random times which is known between the verifier and prover. This way the messages exchanged between an $V$ and $P$ are kept to a minimum. The $P$ creates an attestation response which is sent to the $V$. The $V$ accepts the response only if it is inside the time period for this attestation execution. If that is true then the $V$ checks for authenticity, freshness, and trustworthiness of software measurements corresponding to an accepted software state. This protocol mitigates Denial of Service (DoS) attacks while exhibiting low communication overhead and low network congestion.

\textit{Lightweight Swarm Attestation (LISA)}~\cite{carpent:lisa_ra:2017} aims at investigating methods of attesting a possibly mobile group of provers. A new metric is introduced called Quality of Swarm Attestation (QoSA) that measures the information offered by different collective attestation schemes. LISA provides two practical variations of the collective attestation scheme presented in Figure~\ref{fig:seda-attestation}. LISA's main focus is the construction of the spanning tree.  The first variation includes a synchronous mechanism to aggregate the attestation reports. The second variation uses an asynchronous protocol that forwards the attestation reports directly to the verifier without any previous aggregation.

\textit{Device Attestation Resilient to Physical Attacks (DARPA)}~\cite{ibrahim2016darpa} is a collective attestation scheme that can detect both software-based and physical attacks. The authors propose a heartbeat protocol that is executed at regular time intervals. The heartbeat is propagated throughout the network topology and it proves the presence of neighbouring devices that are members in a network topology. This way an adversary that compromises a device can be detected, under the premise that the adversary needs to take the device offline to attack it. A verifier can then detect if any devices are missing when collecting the heartbeat logs. In addition, this work is can be integrated with other collective attestation schemes such as SEDA or LISA to protect the network topology from an adversary with physical access.

\textit{Secure and Lightweight Attestation of Highly Dynamic and Disruptive Networks (SALAD)}~\cite{kohnhauser2018salad} is a collective attestation scheme for highly dynamic networks. The main difference of this scheme is that instead of creating a spanning tree of the network topology during attestation with the root node being the verifier, the attestation process uses a distributed approach. The idea is to have devices that communicating with other nearby devices to attest the software integrity of each other. Then they exchange the attestation proofs, which also include accumulated attestation proofs of other nearby devices. SALAD protects against DoS attacks when some of the devices prevent an attestation of other devices in the network. In addition, the verifier can receive the attestation result from any participating device, without the need to hold the communication with a particular network device.

\textit{Decentralized Attestation for Device Swarms (DADS)}~\cite{wedaj2019dads} is another scheme that proposes a decentralized collective attestation scheme that is scalable to a large number of embedded devices and supports high mobility networks with a dynamic topology. In this scheme each member device in the network is attested by a device in its vicinity. Each attested device is then assigned to verify other devices. Thus, there is only a need for a central verifier during the initial stages of the attestation process and makes this scheme resilient to single point failures. In addition, DADS offers a local attestation process where a node can locally attest the software integrity of its modules. In case there are any compromised nodes DADS provides a means of graph restructuring the network topology of the attested nodes.

\textit{Scalable Heterogeneous Layered Attestation (SHeLa)}~\cite{rabbani2019shela} is a collective attestation scheme that adds an additional edge layer in between the root verifier and the swarm devices present in the infrastructure network. The edge layer consists of the edge verifier and the swarm nodes present in this layer, which take the role of the prover in this locality. This scheme uses a distributed method of attestation where the verifier can obtain the attestation result from any node in the swarm. Even though this solution is scalable, there are privacy concerns since each edge verifier keeps a log of related information and the root verifier stores information regarding all the layers of the infrastructure.

\textit{Practical Attestation for Highly Dynamic Swarm Topologies (PADS)}~\cite{ambrosin2018pads} discusses a non-interactive attestation scheme for dynamic and unstructured swarm networks. The main feature of this scheme is the reduction of the collective attestation problem to a minimum consensus problem. The provers attest themselves and add their attestation result into an aggregate attestation by sharing their knowledge to adjacent provers that reside in the swarm network.

\textit{Efficient Remote Attestation via Self-Measurement for Unattended Settings (ERASMUS)}~\cite{carpent2018erasmusa,carpent:self_measurement_ra:2019} is a remote attestation scheme that is based on the notion of self-measuring the software code of provers in timed intervals and storing them locally. A verifier can request the measurement history and discern if there is any mobile malicious code present that has tampered with the measurements. The authors discuss the advantages of ERASMUS in comparison with on-demand attestation and how this scheme can be used in the context of collective attestation by integrating with an on-demand attestation scheme such as LISA. ERASMUS can operate both according to a schedule and on-demand when receiving a query from a verifier. The authors claim that ERASMUS is well suited for collective attestation since it supports high mobility of the devices. Even though, this scheme minimizes the prover computations it still uses LISA for the collective attestation inheriting the shortcomings of the scheme.

In all above schemes, the main issue is that the verifiers learn about the structure of the network topology that is being attested. However, there are cases where we would want to keep the structure of the network confidential and provide other means of attesting the nodes in a network. One such case is when a tenant verifier asks a cloud provider that software integrity of her virtualized host is not compromised. The cloud provider would not want to publicize the blueprint of her infrastructure. Therefore, we require to create novel privacy-preserving attestation schemes that can scale to large networks and be resilient in software and physical attacks.

\begin{table}[tb]
\renewcommand\arraystretch{0.9}
\renewcommand{\tabcolsep}{2pt}
\centering 
\caption{Comparison of collective attestation schemes focusing on security aspects}
\begin{tabular}{l c c c c c c c c}
\toprule
\multirow{2}{*}{Scheme}    & Interaction  & Software  & Physical & Attestation  & Cryptographic & \\
 			   & Pattern      & Attacks   & Attacks  & Method       & Primitives & \\
\midrule
SEDA~\cite{asokan2015seda} &  $1-N$       & $\CIRCLE$ & $\CIRCLE$ & Centralized & ECDSA, SHA-1 & \\

\multirow{2}{*}[-2pt]{SANA~\cite{ambrosin2016sana}}
			      & \multirow{2}{*}[-2pt]{$1-N$}    & \multirow{2}{*}[-2pt]{$\CIRCLE$}     & \multirow{2}{*}[-2pt]{$\CIRCLE$}     & \multirow{2}{*}[-2pt]{Centralized}     & \multirow{2}{3.5cm}[-2pt]{\centering Optimistic Aggregate Signature}\\
\addlinespace
\addlinespace
\addlinespace
DARPA~\cite{ibrahim2016darpa} & $1-N$     & $\CIRCLE$     & $\CIRCLE$     & Centralized     & MAC  & \\
\multirow{2}{*}[-2pt]{SeED~\cite{ibrahim:seed:2017}}
			      & \multirow{2}{*}[-2pt]{$1-N$}     & \multirow{2}{*}[-2pt]{$\CIRCLE$}     & \multirow{2}{*}[-2pt]{$\Circle$}     & \multirow{2}{3.5cm}[-2pt]{\centering Centralized}   & \multirow{2}{*}[-2pt]{HMAC-DRBG} \\

\addlinespace
\addlinespace
\addlinespace
LISA~\cite{carpent:lisa_ra:2017}   & $1-N$     & $\CIRCLE$     & $\Circle$     & Centralized     & ECDSA-256, SHA-256  & \\
\multirow{2}{*}[-2pt]{SALAD~\cite{kohnhauser2018salad}}
			      & \multirow{2}{*}[-2pt]{$1-N$}     & \multirow{2}{*}[-2pt]{$\CIRCLE$}     & \multirow{2}{*}[-2pt]{$\CIRCLE$}     & \multirow{2}{*}[-2pt]{Distributed}     & \multirow{2}{3.5cm}[-2pt]{\centering ECDH, SHA-256, HMAC, EdDSA} \\

\addlinespace
\addlinespace
\addlinespace
DADS~\cite{wedaj2019dads}     & $M-N$     & $\CIRCLE$     & $\Circle$     & Decentralized   & ECDSA, HMAC   & \\
\addlinespace
SHeLa~\cite{rabbani2019shela}     & $1-N$     & $\CIRCLE$     & $\Circle$     & Distributed   & SHA-256   & \\
\addlinespace
PADS~\cite{ambrosin2018pads}     & $1-N$     & $\CIRCLE$     & $\Circle$     & Distributed   & HMAC  & \\
\addlinespace
ERASMUS~\cite{carpent2018erasmusa,carpent:self_measurement_ra:2019}     & $1-N$     & $\CIRCLE$     & $\Circle$     & Centralized & HMAC  & \\
\bottomrule
\end{tabular}\\
\footnotesize
\mbox{$\CIRCLE$: supported feature $\LEFTcircle$: partially supported feature  $\Circle$: not supported feature}
\label{tab:sec-comp}
\end{table}


\section{Open Research Problems}
\label{sec:open-res}
The current remote attestation solutions have issues that give room for improvement. In this section, we discuss open research problems and provide avenues of further research in remote attestation schemes. The main focus of this section is about the privacy and confidentiality issues that arise when using remote attestation schemes and possible ways of addressing those issues. 

\subsection{Privacy Issues}
In hardware-based approaches discussed in this report exhibit a number of privacy issues in terms of how the attestation is executed and the amount of information that that the $V$ learns about the structure of the network infrastructure that is attesting. An issue that arises mainly when attesting nodes in a network is the information that the intermediary nodes learn about the attestation result of adjacent nodes and how resilient is the attestation scheme when the node attested is a malicious one and tries to add its own bogus information to relay to other adjacent nodes in the network. In the attestation schemes that adjacent nodes learn the attestation result of the previous nodes this can be an issue in terms of relaying reliable information to the other nodes and finally the $V$ will receive attestation results that will not be correct in case that a malicious node injects its own attestation results. 

Therefore, a method to relay attestation results without the intermediary network nodes to know is beneficial to improve the privacy of an attestation scheme. A number of approaches could be selected to achieve the required privacy level. One such approach involves cryptographic primitives that are privacy-preserving and hide information from potential adversaries. 

Anonymous credential schemes provide a way to construct privacy-preserving attestation schemes for nodes in a  network. DAA is one of the most prevalent cryptographic scheme based on anonymous credentials and group signatures. When we attest multiple platforms in compound proof statements in a network infrastructure, a cryptographic scheme based on anonymous credentials would be a step forward for protecting the privacy of intermediary attestation results which include the software configuration of the device.

\subsection{Confidentiality Issues}

Apart from the privacy issues mentioned above there are a number of issues regarding the confidentiality of attestation schemes in network infrastructures. In collective attestation schemes a malicious prover that is adjacent to a number of other nodes in a network can discover the structure of the network using the attestation results of other nodes that it is collecting. In order to keep the structure of the network secret, there needs to be a method of protecting the blueprint of the network without disclosing its structure. One way is to use signature schemes and zero knowledge proof of knowledge protocols to convince adjacent provers that the structure of the network is known but the actual structure is not disclosed to other provers. Then the verifier can be assured that the structure of the network has been protected. This method make sure that the verifier also does not get to know the structure of the network from the attestation results collected from the network nodes. 

\section{Conclusions}
Attesting network infrastructures is an important avenue for ensuring that nodes in a network are in a legitimate state and are authenticated. There are a number of hardware architectures that have been proposed to further support attestation schemes. A number of these hardware architectures support higher-end hardware and cryptographic primitives for attestation schemes, while other architectures exhibit a more minimal approach supporting lower-end hardware and lightweight cryptographic primitives. We presented a categorization of attestation schemes depending on the type of the attestation each research work used and the cryptographic primitives employed to realize the schemes. This allowed us to compare and contrast the attestation schemes that focus on different aspects on how to create attestation schemes for network infrastructures.  We also identified open research issues in terms of confidentiality and privacy. Finally, we presented possible avenues of resolving the issues in attestation schemes.

This paper shows that there are many works related to attestation architectures and schemes. However, there is still room for improvement in terms of scalability, privacy and confidentiality issues for attestation. We hope this survey acts as a starting point to future research on attestation. 

\section*{Acknowledgement}
This work was supported in full by the \CASCAde. 

\newpage

\bibliography{bibliography}

\begin{thebibliography}{10}

\bibitem{ambrosin:sana:2016}
Moreno Ambrosin, Mauro Conti, Ahmad Ibrahim, Gregory Neven, Ahmad-Reza Sadeghi,
  and Matthias Schunter.
\newblock {SANA: Secure and Scalable Aggregate Network Attestation}.
\newblock In {\em Proceedings of the 2016 ACM SIGSAC Conference on Computer and
  Communications Security}, pages 731--742, 2016.

\bibitem{ambrosin2016sana}
Moreno Ambrosin, Mauro Conti, Ahmad Ibrahim, Gregory Neven, Ahmad-Reza Sadeghi,
  and Matthias Schunter.
\newblock {SANA: secure and scalable aggregate network attestation}.
\newblock In {\em Proceedings of the 2016 ACM SIGSAC Conference on Computer and
  Communications Security}, pages 731--742. ACM, 2016.

\bibitem{ambrosin:pads:2018}
Moreno Ambrosin, Mauro Conti, Riccardo Lazzeretti, Md~Masoom Rabbani, and
  Silvio Ranise.
\newblock {PADS: Practical Attestation for Highly Dynamic Swarm Topologies}.
\newblock 2018 International Workshop on Secure Internet of Things (SIoT),
  pages 18--27, 2018.

\bibitem{ambrosin2018pads}
Moreno Ambrosin, Mauro Conti, Riccardo Lazzeretti, Md~Masoom Rabbani, and
  Silvio Ranise.
\newblock {PADS: Practical Attestation for Highly Dynamic Swarm Topologies}.
\newblock 2018 International Workshop on Secure Internet of Things (SIoT),
  pages 18--27, 2018.

\bibitem{anati:intel_sgxra:2013}
I~Anati, S~Gueron, Simon~P Johnson, and Vincent~R Scarlata.
\newblock {Innovative technology for CPU based attestation and sealing}.
\newblock In {\em Proceedings of the 2nd International Workshop on Hardware and
  Architectural Support for Security and Privacy}, 2013.

\bibitem{armknecht:software_attestation:2013}
Frederik Armknecht, Ahmad-Reza Sadeghi, Steffen Schulz, and Christian
  Wachsmann.
\newblock {A Security Framework for the Analysis and Design of Software
  Attestation}.
\newblock In {\em Proceedings of the 2013 ACM SIGSAC Conference on Computer \&
  Communications Security}, CCS '13, pages 1--12, New York, NY, USA, 2013.
  Association for Computing Machinery.

\bibitem{asokan2015seda}
Nadarajah Asokan, Ferdinand Brasser, Ahmad Ibrahim, Ahmad-Reza Sadeghi,
  Matthias Schunter, Gene Tsudik, and Christian Wachsmann.
\newblock {SEDA: Scalable embedded device attestation}.
\newblock In {\em Proceedings of the 22nd ACM SIGSAC Conference on Computer and
  Communications Security}, pages 964--975. ACM, 2015.

\bibitem{bays:2015}
Leonardo~Richter Bays, Rodrigo~Ruas Oliveira, Marinho~Pilla Barcellos,
  Luciano~Paschoal Gaspary, and Edmundo Roberto~Mauro Madeira.
\newblock {Virtual network security: threats, countermeasures, and challenges}.
\newblock {\em Journal of Internet Services and Applications}, 6(1):1, 2015.

\bibitem{bernhard2013aat}
D.~Bernhard, G.~Fuchsbauer, E.~Ghadafi, N.~P. Smart, and B.~Warinschi.
\newblock Anonymous attestation with user-controlled linkability.
\newblock {\em International Journal of Information Security}, 12(3):219--249,
  Jun 2013.

\bibitem{braser:tytan:2015}
Ferdinand Brasser, Brahim~El Mahjoub, Ahmad-Reza Sadeghi, Christian Wachsmann,
  and Patrick Koeberl.
\newblock {TyTAN: Tiny Trust Anchor for Tiny Devices}.
\newblock 52nd ACM/EDAC/IEEE Design Automation Conference (DAC), pages 1--6,
  2015.

\bibitem{rasmussen2016DAC}
Ferdinand Brasser, Kasper Rasmussen, Ahmad-Reza Sadeghi, and Gene Tsudik.
\newblock Remote attestation for low-end embedded devices: the prover's
  perspective.
\newblock In {\em Design Automation Conference {(DAC)}}, June 2016.

\bibitem{brickell2010epidbp}
E.~{Brickell} and J.~{Li}.
\newblock Enhanced privacy id from bilinear pairing for hardware authentication
  and attestation.
\newblock In {\em 2010 IEEE Second International Conference on Social
  Computing}, pages 768--775, Aug 2010.

\bibitem{brickell2004direct}
Ernie Brickell, Jan Camenisch, and Liqun Chen.
\newblock Direct anonymous attestation.
\newblock In {\em Proceedings of the 11th ACM conference on Computer and
  communications security}, pages 132--145. ACM, 2004.

\bibitem{brickell:2008}
Ernie Brickell, Liqun Chen, and Jiangtao Li.
\newblock {A New Direct Anonymous Attestation Scheme from Bilinear Maps}.
\newblock pages 166--178, 2008.

\bibitem{brickell2008daabm}
Ernie Brickell, Liqun Chen, and Jiangtao Li.
\newblock A new direct anonymous attestation scheme from bilinear maps.
\newblock In Peter Lipp, Ahmad-Reza Sadeghi, and Klaus-Michael Koch, editors,
  {\em Trusted Computing - Challenges and Applications}, pages 166--178,
  Berlin, Heidelberg, 2008. Springer Berlin Heidelberg.

\bibitem{brickell2009daanotions}
Ernie Brickell, Liqun Chen, and Jiangtao Li.
\newblock Simplified security notions of direct anonymous attestation and a
  concrete scheme from pairings.
\newblock {\em International Journal of Information Security}, 8(5):315--330,
  Oct 2009.

\bibitem{brickell2007enhanced}
Ernie Brickell and Jiangtao Li.
\newblock Enhanced privacy id: A direct anonymous attestation scheme with
  enhanced revocation capabilities.
\newblock In {\em Proceedings of the 2007 ACM workshop on Privacy in electronic
  society}, pages 21--30. ACM, 2007.

\bibitem{brickell2010daapair}
Ernie Brickell and Jiangtao Li.
\newblock A pairing-based daa scheme further reducing tpm resources.
\newblock In Alessandro Acquisti, Sean~W. Smith, and Ahmad-Reza Sadeghi,
  editors, {\em Trust and Trustworthy Computing}, pages 181--195, Berlin,
  Heidelberg, 2010. Springer Berlin Heidelberg.

\bibitem{camenisch2017bind}
J.~{Camenisch}, L.~{Chen}, M.~{Drijvers}, A.~{Lehmann}, D.~{Novick}, and
  R.~{Urian}.
\newblock One tpm to bind them all: Fixing tpm 2.0 for provably secure
  anonymous attestation.
\newblock In {\em 2017 IEEE Symposium on Security and Privacy (SP)}, pages
  901--920, May 2017.

\bibitem{camenisch2006aac}
Jan Camenisch.
\newblock {Protecting (Anonymous) Credentials with the Trusted Computing
  Group's TPM V1.2}.
\newblock In Simone Fischer-H{\"u}bner, Kai Rannenberg, Louise Yngstrom, and
  Stefan Lindskog, editors, {\em Security and Privacy in Dynamic Environments},
  pages 135--147, Boston, MA, 2006. Springer US.

\bibitem{camenisch2016aa}
Jan Camenisch, Manu Drijvers, and Anja Lehmann.
\newblock Anonymous attestation using the strong diffie hellman assumption
  revisited.
\newblock In Michael Franz and Panos Papadimitratos, editors, {\em Trust and
  Trustworthy Computing}, pages 1--20, Cham, 2016. Springer International
  Publishing.

\bibitem{camenisch2016ucdaa}
Jan Camenisch, Manu Drijvers, and Anja Lehmann.
\newblock {Universally Composable Direct Anonymous Attestation}.
\newblock In Chen-Mou Cheng, Kai-Min Chung, Giuseppe Persiano, and Bo-Yin Yang,
  editors, {\em Public-Key Cryptography -- PKC 2016}, pages 234--264, Berlin,
  Heidelberg, 2016. Springer Berlin Heidelberg.

\bibitem{carpent:lisa_ra:2017}
Xavier Carpent, Karim ElDefrawy, Norrathep Rattanavipanon, and Gene Tsudik.
\newblock {Lightweight Swarm Attestation: A Tale of Two LISA-s}.
\newblock In {\em Proceedings of the 2017 ACM on Asia Conference on Computer
  and Communications Security}, pages 86--100, 2017.

\bibitem{carpent2018erasmusa}
Xavier Carpent, Norrathep Rattanavipanon, and Gene Tsudik.
\newblock {ERASMUS: Efficient Remote Attestation via Self-Measurement for
  Unattended Settings}.
\newblock volume 2018-January of {\em 2018 Design, Automation \& Test in Europe
  Conference \& Exhibition (DATE)}, pages 1191--1194, 2018.

\bibitem{carpent:self_measurement_ra:2019}
Xavier Carpent, Norrathep Rattanavipanon, and Gene Tsudik.
\newblock {Remote Attestation via Self-Measurement}.
\newblock {\em ACM Transactions on Design Automation of Electronic Systems
  (TODAES)}, 24(1):11, 2019.

\bibitem{chasaki:attacks_networkinfra:2011}
Danai Chasaki, Qiang Wu, and Tilman Wolf.
\newblock {Attacks on Network Infrastructure}.
\newblock volume~1 of {\em 2011 Proceedings of 20th International Conference on
  Computer Communications and Networks (ICCCN)}, pages 1--8, 2011.

\bibitem{chen:sgx_opera_ra:2019}
Guoxing Chen, Yinqian Zhang, and Ten-Hwang Lai.
\newblock {OPERA: Open Remote Attestation for Intel's Secure Enclaves}.
\newblock In {\em Proceedings of the 2019 ACM SIGSAC Conference on Computer and
  Communications Security}, pages 2317--2331, 2019.

\bibitem{chen2010daaless}
Liqun Chen.
\newblock A daa scheme requiring less tpm resources.
\newblock In Feng Bao, Moti Yung, Dongdai Lin, and Jiwu Jing, editors, {\em
  Information Security and Cryptology}, pages 350--365, Berlin, Heidelberg,
  2010. Springer Berlin Heidelberg.

\bibitem{chen2010daabatch}
Liqun Chen.
\newblock A daa scheme using batch proof and verification.
\newblock In Alessandro Acquisti, Sean~W. Smith, and Ahmad-Reza Sadeghi,
  editors, {\em Trust and Trustworthy Computing}, pages 166--180, Berlin,
  Heidelberg, 2010. Springer Berlin Heidelberg.

\bibitem{chen:2008}
Liqun Chen, Hans L{\"o}hr, Mark Manulis, and Ahmad-Reza Sadeghi.
\newblock {Property-Based Attestation without a Trusted Third Party}.
\newblock pages 31--46, 2008.

\bibitem{chen2010daaeff}
Liqun Chen, Dan Page, and Nigel~P. Smart.
\newblock On the design and implementation of an efficient daa scheme.
\newblock In Dieter Gollmann, Jean-Louis Lanet, and Julien Iguchi-Cartigny,
  editors, {\em Smart Card Research and Advanced Application}, pages 223--237,
  Berlin, Heidelberg, 2010. Springer Berlin Heidelberg.

\bibitem{chen2015daaa}
Liqun Chen and Rainer Urian.
\newblock Daa-a: Direct anonymous attestation with attributes.
\newblock In Mauro Conti, Matthias Schunter, and Ioannis Askoxylakis, editors,
  {\em Trust and Trustworthy Computing}, pages 228--245, Cham, 2015. Springer
  International Publishing.

\bibitem{chen:priv_ca:2010}
Liqun Chen and Bogdan Warinschi.
\newblock {Security of the TCG Privacy-CA Solution}.
\newblock 2010 IEEE/IFIP International Conference on Embedded and Ubiquitous
  Computing, pages 609--616, 2010.

\bibitem{cisa:alert:2018}
CISA.
\newblock {Russian State-Sponsored Cyber Actors Targeting Network
  Infrastructure Devices}, 2018.

\bibitem{cisa:2020}
CISA.
\newblock {Critical Infrastructure Sectors}, 2020.

\bibitem{coker:principles_ra:2011}
George Coker, Joshua Guttman, Peter Loscocco, Amy Herzog, Jonathan Millen,
  Brian O'Hanlon, John Ramsdell, Ariel Segall, Justin Sheehy, and Brian
  Sniffen.
\newblock {Principles of Remote Attestation}.
\newblock {\em International Journal of Information Security}, 10(2):63 81,
  2011.

\bibitem{costan:sanctum:2016}
Victor Costan, Ilia~A Lebedev, and Srinivas Devadas.
\newblock {Sanctum - Minimal Hardware Extensions for Strong Software
  Isolation.}
\newblock In {\em 25th \$\textbackslash\{\$USENIX\$\textbackslash\}\$ Security
  Symposium (\$\textbackslash\{\$USENIX\$\textbackslash\}\$ Security 16)},
  USENIX Security Symposium, pages 857---874, 2016.

\bibitem{datta:logic_trustedcomputing:2009}
Anupam Datta, Jason Franklin, Deepak Garg, and Dilsun Kaynar.
\newblock {A Logic of Secure Systems and its Application to Trusted Computing}.
\newblock 2009 30th IEEE Symposium on Security and Privacy, pages 221--236,
  2009.

\bibitem{defrawy:2012}
Karim~El Defrawy, Aur\'elien Francillon, Daniele Perito, and Gene Tsudik.
\newblock {SMART: Secure and Minimal Architecture for (Establishing a Dynamic)
  Root of Trust}.
\newblock In {\em 19th Annual NDSS Symposium}, 2012.

\bibitem{enisa:smart_grid:2012}
ENISA.
\newblock {Smart Grid Security: Annex II. Security Aspects of Smart Grid}.
\newblock Technical report, ENISA, 2012.

\bibitem{francillon2014minimalist}
Aur{\'e}lien Francillon, Quan Nguyen, Kasper~B Rasmussen, and Gene Tsudik.
\newblock {A Minimalist Approach to Remote Attestation}.
\newblock {\em 2014 Design, Automation \& Test in Europe Conference \&
  Exhibition (DATE)}, pages 1--6, 2014.

\bibitem{gassend:puf:2002}
Blaise Gassend, Dwaine Clarke, Marten~van Dijk, and Srinivas Devadas.
\newblock {Silicon Physical Random Functions}.
\newblock In {\em Proceedings of the 9th ACM Conference on Computer and
  Communications Security}, pages 148--160, 2002.

\bibitem{hes2007daaemb}
He~Ge and Stephen~R. Tate.
\newblock A direct anonymous attestation scheme for embedded devices.
\newblock In Tatsuaki Okamoto and Xiaoyun Wang, editors, {\em Public Key
  Cryptography -- PKC 2007}, pages 16--30, Berlin, Heidelberg, 2007. Springer
  Berlin Heidelberg.

\bibitem{tcg:2005}
Trusted~Computing Group.
\newblock {Trusted Computing Platform Alliance (TCPA) Main Specification
  Version 1.1b}.
\newblock 2005.

\bibitem{tpm1.2}
Trusted~Computing Group.
\newblock {TPM Main Specification Level 2 Version 1.2, Revision 116}, 2011.
\newblock Last accessed: 21 November 2019.

\bibitem{tcg:tpm2:architecture}
Trusted~Computing Group.
\newblock {Trusted Platform Module Library Part 1: Architecture Family 2.0
  Level 00 Revision 01.59}.
\newblock Technical report, 2019.

\bibitem{ibrahim:us-aid:2018}
Ahmad Ibrahim, Ahmad-Reza Sadeghi, and Gene Tsudik.
\newblock {US-AID: Unattended Scalable Attestation of IoT Devices}.
\newblock IEEE 37th Symposium on Reliable Distributed Systems (SRDS), pages
  21--30, 2018.

\bibitem{ibrahim:seed:2017}
Ahmad Ibrahim, Ahmad-Reza Sadeghi, and Shaza Zeitouni.
\newblock {SeED: Secure Non-Interactive Attestation for Embedded Devices}.
\newblock In {\em Proceedings of the 10th ACM Conference on Security and
  Privacy in Wireless and Mobile Networks}, pages 64--74, 2017.

\bibitem{ibrahim2016darpa}
Ahmad Ibrahim, Ahmad~Reza Sadeghi, Shaza Zeitouni, and Gene Tsudik.
\newblock {DARPA: Device attestation resilient to physical attacks}.
\newblock In {\em WiSec 2016 - Proceedings of the 9th ACM Conference on
  Security and Privacy in Wireless and Mobile Networks}, pages 171--182, TU
  Darmstadt, Germany, 2016. Association for Computing Machinery, Inc.

\bibitem{johnson:sgx_epid_prov_attest:2016}
Simon Johnson, Vinnie Scarlata, Carlos Rozas, Ernie Brickell, and Frank Mckeen.
\newblock {Intel Software Guard Extensions: EPID Provisioning and Attestation
  Services}.
\newblock Technical report, 2016.

\bibitem{kassem2019daalat}
Nada~El Kassem, Liqun Chen, Rachid~El Bansarkhani, Ali~El Kaafarani, Jan
  Camenisch, Patrick Hough, Paulo Martins, and Leonel Sousa.
\newblock More efficient, provably-secure direct anonymous attestation from
  lattices.
\newblock {\em Future Generation Computer Systems}, 99:425 -- 458, 2019.

\bibitem{koeberl:trustlite:2014}
Patrick Koeberl, Steffen Schulz, Ahmad-Reza Sadeghi, and Vijay Varadharajan.
\newblock {TrustLite: a security architecture for tiny embedded devices}.
\newblock In {\em Proceedings of the Ninth European Conference on Computer
  Systems}, pages 1--14, 2014.

\bibitem{kohnhauser2018salad}
Florian Kohnh{\"{a}}user, Niklas B{\"{u}}scher, and Stefan Katzenbeisser.
\newblock {SALAD: Secure and lightweight attestation of highly dynamic and
  disruptive networks}.
\newblock In {\em ASIACCS 2018 - Proceedings of the 2018 ACM Asia Conference on
  Computer and Communications Security}, pages 329--342, Security Engineering
  Group, TU Darmstadt, Germany, 2018. Association for Computing Machinery, Inc.

\bibitem{kong:pufatt_ra:2014}
J.~{Kong}, F.~{Koushanfar}, P.~K. {Pendyala}, A.~{Sadeghi}, and C.~{Wachsmann}.
\newblock Pufatt: Embedded platform attestation based on novel processor-based
  pufs.
\newblock In {\em 2014 51st ACM/EDAC/IEEE Design Automation Conference (DAC)},
  pages 1--6, June 2014.

\bibitem{kreutz:sdn_survey:2014}
Diego Kreutz, Fernando M~V Ramos, Paulo~Esteves Verissimo, Christian~Esteve
  Rothenberg, Siamak Azodolmolky, and Steve Uhlig.
\newblock {Software-Defined Networking: A Comprehensive Survey}.
\newblock {\em Proceedings of the IEEE}, 103(1):14--76, 2014.

\bibitem{lara:openflow_survey:2014}
Adrian Lara, Anisha Kolasani, and Byrav Ramamurthy.
\newblock {Network Innovation using OpenFlow: A Survey}.
\newblock {\em IEEE Communications Surveys \& Tutorials}, 16(1):493--512, 2014.

\bibitem{lebedev:sanctum_ra:2018}
Ilia Lebedev, Kyle Hogan, and Srinivas Devadas.
\newblock {Invited Paper: Secure Boot and Remote Attestation in the Sanctum
  Processor}.
\newblock 2018 IEEE 31st Computer Security Foundations Symposium (CSF), pages
  46--60, 2018.

\bibitem{li:framework_softwareattestation:2014}
Li~Li, Hong Hu, Jun Sun, Yang Liu, and Jin~Song Dong.
\newblock {Practical Analysis Framework for Software-Based Attestation Scheme}.
\newblock In {\em Proceedings of the 16th International Conference on Formal
  Engineering Methods (ICFEM’14)}, pages 284--299, 2014.

\bibitem{mckeen:intel_sgxra:2013}
Frank McKeen, Alex, Ilya rovich, Alex Berenzon, Carlos~V. Rozas, Hisham Shafi,
  Vedvyas Shanbhogue, and Uday~R. Savagaonkar.
\newblock {Innovative Instructions and Software Model for Isolated Execution}.
\newblock In {\em Proceedings of the 2nd International Workshop on Hardware and
  Architectural Support for Security and Privacy}, 2013.

\bibitem{meng2018dynamic}
Wenjuan Meng, Tao Jiang, and Jianhua Ge.
\newblock Dynamic swarm attestation with malicious devices identification.
\newblock {\em IEEE Access}, 6:50003--50013, 2018.

\bibitem{mijumbi:nfv_survey:2015}
Rashid Mijumbi, Joan Serrat, Juan-Luis Gorricho, Niels Bouten, Filip~De Turck,
  and Raouf Boutaba.
\newblock {Network Function Virtualization: State-of-the-Art and Research
  Challenges}.
\newblock {\em IEEE Communications Surveys \& Tutorials}, 18(1):236--262, 2016.

\bibitem{rabbani2019shela}
Md~Masoom Rabbani, Jo~Vliegen, Jori Winderickx, Mauro Conti, and Nele Mentens.
\newblock {SHeLa: Scalable Heterogeneous Layered Attestation}.
\newblock {\em IEEE Internet of Things Journal}, 2019.

\bibitem{sadeghi2011short}
Ahmad-Reza Sadeghi, Steffen Schulz, and Christian Wachsmann.
\newblock Lightweight remote attestation using physical functions.
\newblock {\em WiSec'11}, 2011.

\bibitem{Sadeghi:2004:PAC:1065907.1066038}
Ahmad-Reza Sadeghi and Christian St\"{u}ble.
\newblock Property-based attestation for computing platforms: Caring about
  properties, not mechanisms.
\newblock In {\em Proceedings of the 2004 Workshop on New Security Paradigms},
  NSPW '04, pages 67--77, New York, NY, USA, 2004. ACM.

\bibitem{safford2005trusted}
David Safford and Mimi Zohar.
\newblock {A trusted Linux client (TLC)}.
\newblock {\em Technical Paper, IBM Research}, 1:1--9, 2005.

\bibitem{Sailer:2004wa}
Reiner Sailer, Xiaolan Zhang, and Trent Jaeger.
\newblock {Design and Implementation of a TCG-based Integrity Measurement
  Architecture}.
\newblock volume~13 of {\em USENIX Security symposium}, pages 223--238, 2004.

\bibitem{sakhini:iot_smartgrid:2019}
Jacob Sakhnini, Hadis Karimipour, Ali Dehghantanha, Reza~M Parizi, and Gautam
  Srivastava.
\newblock {Security aspects of Internet of Things aided smart grids: A
  bibliometric survey}.
\newblock {\em Internet of Things}, 2019.

\bibitem{scarlata:sgx_attest_prim:2018}
Vinnie Scarlata, Simon Johnson, James Beaney, and Piotr Zmijewski.
\newblock {Supporting Third Party Attestation for Intel SGX with Intel Data
  Center Attestation Primitives}.
\newblock Technical report, 2018.

\bibitem{tcg:tpm2}
Michael Scott.
\newblock {Trusted Platform Module Library Specification, Family 2.0, Level 00,
  Revision 01.38}.
\newblock Accessed April 4, 2019.

\bibitem{steiner:2016}
Rodrigo~Viera Steiner and Emil Lupu.
\newblock {Attestation in Wireless Sensor Networks: A Survey}.
\newblock {\em ACM Computing Surveys (CSUR)}, 49:51, 2016.

\bibitem{tan:tpm_wsn:2015}
Hailun Tan, Wen Hu, and Sanjay Jha.
\newblock {A remote attestation protocol with Trusted Platform Modules (TPMs)
  in wireless sensor networks.}
\newblock {\em Security and Communication Networks}, 8(13):2171--2188, 2015.

\bibitem{wang:intel_sgx_ra:2017}
Juan Wang, Zhi Hong, Yuhan Zhang, and Yier Jin.
\newblock {Enabling Security-Enhanced Attestation With Intel SGX for Remote
  Terminal and IoT}.
\newblock {\em IEEE Transactions on Computer-Aided Design of Integrated
  Circuits and Systems}, 37(1):88--96, 2017.

\bibitem{wedaj2019dads}
Samuel Wedaj, Kolin Paul, and Vinay~J Ribeiro.
\newblock {DADS: Decentralized Attestation for Device Swarms}.
\newblock {\em ACM Transactions on Privacy and Security (TOPS)}, 22(3):19,
  2019.

\bibitem{xiaofeng2008direct}
Chen Xiaofeng and Feng Dengguo.
\newblock {Direct Anonymous Attestation for Next Generation TPM}.
\newblock {\em Journal of Computers}, 3(12):43--50, 2008.

\bibitem{yang2018direct}
Kang Yang, Liqun Chen, Zhenfeng Zhang, Chris Newton, Bo~Yang, and Li~Xi.
\newblock {Direct Anonymous Attestation with Optimal TPM Signing Efficiency.}
\newblock {\em IACR Cryptology ePrint Archive}, 2018:1128, 2018.

\end{thebibliography}
\bibliographystyle{plain}

\end{document}